# Matter wave propagation above a step potential within the cubic-nonlinear Schrödinger equation

H.A. Ishkhanyan[1,2], A.M. Manukyan[2], and A.M. Ishkhanyan[2]

[1]*Moscow Institute of Physics and Technology, 141700 Dolgoprudny, Russia*
[2]*Institute for Physical Research of NAS of Armenia, 0203 Ashtarak-2, Armenia*

**Abstract.** We analyze the matter wave transmission above a step potential within the framework of the cubic-nonlinear Schrödinger equation. We present a comprehensive analysis of the corresponding stationary problem based on an exact second-order nonlinear differential equation for the probability density. The exact solution of the problem in terms of the Jacobi elliptic *sn*-function is presented and analyzed. Qualitatively distinct types of wave propagation picture are classified depending on the input parameters of the system. Analyzing the 2D space of involved dimensionless parameters, the nonlinearity and the reflecting potential's height/depth given in the units of the chemical potential, we show that the region of the parameters that does not sustain restricted solutions is given by a closed curve consisting of a segment of an elliptic curve and two line intervals. We show that there exists a specific singular point, belonging to the elliptic curve, which causes a jump from one evolution scenario to another one. The position of this point is determined and the peculiarities of the evolution scenarios (oscillatory, non-oscillatory and diverging) for all the allowed regions of involved parameters are described and analyzed in detail.

**PACS numbers:** 03.75.-b Matter waves, 67.85.Hj Bose-Einstein condensates in optical potentials, 03.75.Lm Tunneling, Josephson effect, Bose-Einstein condensates in periodic potentials, solitons, vortices, and topological excitations, 03.65.Ge Solutions of wave equations: bound states

**Keywords:** Gross-Pitaevskii equation, above-barrier reflection, resonance transmission, quantum tunneling

The nonlinear Schrödinger equation refers to different nonlinear generalizations of the Schrödinger equation [1] which are encountered in many fields of contemporary physics research, e.g., in fluid mechanics, nonlinear optics, plasma physics, statistical physics, field theory, etc. (see, e.g., [2]). It is an example of a universal nonlinear model that is applicable to many particular phenomena in various nonlinear physical systems. The *cubic-nonlinear* version of the nonlinear Schrödinger equation, being probably the simplest polynomial generalization of the linear Schrödinger equation, allows understanding of general mechanisms underlying nonlinear dynamics in many cases in hydrodynamics, nonlinear optics, nonlinear acoustics, quantum condensates, heat pulses in solids and various other nonlinear phenomena. This equation, also known as Gross-Pitaevskii equation [3], has been attracting much interest in recent years because it describes the dynamics of Bose-Einstein condensates in the mean-field approximation at ultralow temperatures [4].

Among many aspects discussed in the context of Bose-condensates' dynamics within the framework of the Gross-Pitaevskii approximation, the quantum transport across and reflection from potential barriers/wells is of interest since these phenomena are pure quantum effects and the nonlinearity offers a different discussion of the basic concepts of quantum mechanics. It is therefore of wide interest the characterization of the complete set of solutions for the matter-wave transport across model potentials as a function of the nonlinearity parameter. Several such models have been recently discussed including the step-potential [5], the rectangular barrier [6], the delta, double-delta and Kronig-Penney configurations [7], the Rosen-Morse potential [8-9], etc. It should be noted, however, that the treatment presented in these papers is either applicable to a restricted variation range of involved parameters or the whole parameter range is not discussed in detail. In the present paper, we present a comprehensive analysis of allowed stationary solutions for the matter-wave transport above a step potential. The analysis is based on the exact solution of the problem and, hence, is applicable for the whole variation range of the parameters of the system. We analyze the two-dimensional space of involved dimensionless parameters, the nonlinearity and potential's height/depth, and divide it into two regions where qualitatively distinct types of wave propagation occur. It turns out that unlike the linear case the behavior of the system may be non-oscillatory and, moreover, the solution may diverge. We show that the region of parameters for which the solution is not restricted is bounded by two line intervals and a segment of an elliptic curve. We determine the explicit form of this curve; find the position of the most singular point belonging to it, which stands for the change from one evolution scenario to another one. Furthermore, we present and analyze in detail the corresponding evolution scenarios (oscillatory, non-oscillatory and diverging) for all the regions of the parameter space.

The cubic-nonlinear Schrödinger equation in the one-dimensional case is written as

$$i\hbar \frac{\partial \Psi}{\partial t} = -\frac{\hbar^2}{2m}\frac{\partial^2 \Psi}{\partial x^2} + (V(x) + g|\Psi|^2)\Psi. \qquad (1)$$

Here, the nonlinearity parameter $g$ determining the mean-field self-interaction is given through the $s$-wave scattering length $a_s$ for binary elastic collisions of two interacting bosons of mass $m$ as $g = 4\pi\hbar^2 a_s/m$, and $V(x)$ is the external field's potential. In the present paper, we consider a reflecting step potential given as: $V(x) = 0$ if $x < 0$ and $V(x) = V_0$ if $x \geq 0$. The nonlinearity parameter may be positive or negative, corresponding to repulsive or attractive interactions, respectively. For instance, in Bose-Einstein condensates of dilute atomic gases [4], it is possible to alter the s-wave scattering length and hence the nonlinearity of the nonlinear Schrödinger equation using the Feshbach resonances. Near a Feshbach resonance, the scattering length is a sensitive function of the applied uniform magnetic field. By altering the field, the effect of the nonlinearity can substantially be controlled. Since the nonlinearity alters the effective potential experienced by the system it is important to analyze the solution as a function of the interaction strength and the parameters of the potential. Below we present the analysis for the step-potential based on the exact solution of Eq. (1). Though the repulsive interaction (positive nonlinearity) is more common in experimental condensates, our treatment is applicable to the case of attractive interaction as well.

Applying the ansatz $\Psi(x,t) = \exp(-i\mu t/\hbar)\psi(x)$, where $\mu$ is the chemical potential for a conserved number of particles, Eq. (1) is reduced to the following stationary version

$$-\frac{1}{2}\frac{d^2\psi}{dx^2} + (-\mu + V(x) + g|\psi|^2)\psi = 0 \qquad (2)$$

(we use units such that $\hbar = m = 1$). Applying further the transformation $\psi(x) = \sqrt{p}e^{i\theta(x)}$ and separating the real and imaginary terms, one gets the following equations for the probability density $p = |\psi(x)|^2$ and phase $\theta(x)$:

$$-\frac{1}{2}\left(\frac{p''p}{2} - \frac{(p')^2}{4}\right) + (-\mu + V + gp)p^2 + (p\theta')^2 = 0, \qquad (3)$$

$$\frac{p'\theta'}{\sqrt{p}} + \sqrt{p}\theta'' = 0, \qquad (4)$$

where the prime denotes differentiation with respect to $x$. The second equation is readily integrated with the result $p\theta' = J_0 = \text{const}$, which implies that the system is flow-conserving ($p\theta' = i(\psi'\bar{\psi} - \psi\bar{\psi}')/2$) and, hence, the last term of Eq. (3) is constant everywhere. With this observation, Eq. (3) for probability density is rewritten as

$$\frac{p''p}{4} - \frac{(p')^2}{8} = (-\mu + V + gp)p^2 + J_0^2. \qquad (5)$$

In the present paper, we look for a solution with traveling-wave-like asymptotic behavior $\psi(-\infty) \sim \sqrt{p_0}e^{ikx}$, $k = \sqrt{2(\mu - gp_0)}$, hence, the function $p(x)$ obeys the initial conditions

$$p(-\infty) = p_0, \quad p'(-\infty) = 0, \quad p''(-\infty) = 0. \qquad (6)$$

Since it follows from the Gross-Pitaevskii equation that the normalization of the wave function can always be incorporated into the definition of the nonlinearity parameter $g$, without loss of generality we choose the normalization $p_0 = 1$ so that we assume $p(-\infty) = 1$. Since $V(-\infty) = 0$, for the flow $J_0$ we get $J_0^2 = \mu - g = k^2/2$.

Consider the region $x \geq 0$, where $V = V_0$. If $p'$ is not identically zero everywhere, that is if $p \neq p_0 = 1$, multiplying Eq. (5) by $p'/p^2$ leads to exact integration with the result

$$\frac{1}{8}\left(\frac{dp}{dx}\right)^2 = \left((-\mu + V_0)p^2 + \frac{g}{2}p^3\right) + C_1 p - J_0^2. \tag{7}$$

The initial conditions at $x = 0$ read $p(0) = 1$ and $p'(0) = 0$, hence, for the integration constant $C_1$ we get

$$C_1 = 2\mu - V_0 - \frac{3g}{2}. \tag{8}$$

Eq. (7) is then rewritten as

$$\left(\frac{dp}{dx}\right)^2 = 4(1-p)\left(-k^2 + [k_a^2 + g(1-p)]p\right), \tag{9}$$

where $k_a = \sqrt{2(\mu - g - V_0)}$. We note that if $g \neq 0$ the right-hand side of this equation is a cubic polynomial in $p$ while in the linear case $g = 0$ it becomes a quadratic one. The corresponding equation,

$$(p_L')^2 = 4(1 - p_L)\left(-k_L^2 + k_{La}^2 p_L\right), \tag{10}$$

then reproduces the known result [1]:

$$p_L = 1 + \left(\frac{k_L^2}{k_{La}^2} - 1\right)\left(\sin(k_{La}x)\right)^2, \tag{11}$$

where $k_L = \sqrt{2\mu}$ and $k_{La} = \sqrt{2(\mu - V_0)}$.

In the general case, the transformation $p = 1 + e_1(u(bx))^2$ with $b = \sqrt{g e_2}$ reduces Eq. (9) to the equation obeyed by the Jacobi elliptic $sn$ function [10]:

$$u'^2 = (1 - u^2)(1 - mu^2) \tag{12}$$

with $\quad m = \dfrac{e_1}{e_2}, \quad e_{1,2} = \dfrac{k_a^2 - g \mp \sqrt{(k_a^2 + g)^2 - 4gk^2}}{2g}. \tag{13}$

Thus, the solution for $x \geq 0$ is written as

$$p = 1 + e_1 \left( sn\left[ \sqrt{g e_2} x, \frac{e_1}{e_2} \right] \right)^2. \tag{14}$$

The structure of this solution suggests several evolution scenarios depending on the sign of discriminant $D = (k_a^2 + g)^2 - 4gk^2$. Depending on whether the elliptic parameters $e_1$ and $e_2$ are real or complex (that is if $D \geq 0$ or $D < 0$), equal or not, the characteristics of the solution essentially change. As it was already mentioned above, unlike the linear case when the asymptote of the solution at $x \to +\infty$ is always oscillatory (indicating the interference of incident and reflected waves in that case), the solution in the nonlinear case under consideration may be non-oscillatory and, moreover, diverging. In order to classify the possible types of solutions, we now analyze the two-dimensional space of dimensionless input parameters $\{ g/\mu, V_0/\mu \}$. Here, we should note that the general restrictions imposed on the parameters in order that the matter-wave transport be *above* the barrier/well the wave numbers $k$ and $k_a$ should be real. This implies that $g \leq \mu$ for $V_0 \leq 0$ (i.e., for ascending potential barriers since we assume that the matter wave moves from right to left) and $g < \mu - V_0$ for $V_0 > 0$ (i.e., for descending potential barriers).

If $D < 0$ the elliptic parameters $e_1$ and $e_2$ are complex and the solution diverges at $x \to +\infty$. This happens if the potential height $V_0$ and the nonlinearity parameter $g$ belong to the region marked by the darker grey in Fig. 1 (note that in this figure the parameters are given in the units of the chemical potential $\mu$). The borders of this region are defined by the closed curve $C_{ASBA}$,

$$C_{ASBA} : \left\{ V_0 = \frac{1}{2} \left( \sqrt{g} - \sqrt{2(\mu - g)} \right)^2 \,\&\, g = \mu - V_0 \right\}, \tag{15}$$

composed of a segment of the elliptic curve $C_{ASB}$ and the interval $[A, B]$ of the line $g = \mu - V_0$; $A = \{ \mu/9, 8\mu/9 \}$.

On the elliptic curve $C_{ASB}$ defined by the equation

$$V_0 = \frac{1}{2} \left( \sqrt{g} - \sqrt{2(\mu - g)} \right)^2 \tag{16}$$

the discriminant vanishes, $D = 0$, and the two elliptic parameters coalesce: $e_{1,2} = (k_a^2 - g)/(2g)$. Note that this curve is located in the first quadrant so that this situation is possible only if the nonlinearity parameter is positive, $g > 0$. Though in this case the parameters are real, however, the square root in the argument of *sn*-function of Eq. (14) suggests that the solution will exponentially grow if $e_2 < 0$. This situation occurs on the upper branch *SA* of the curve $C_{ASB}$ (Fig. 1, dashed line *SA*).

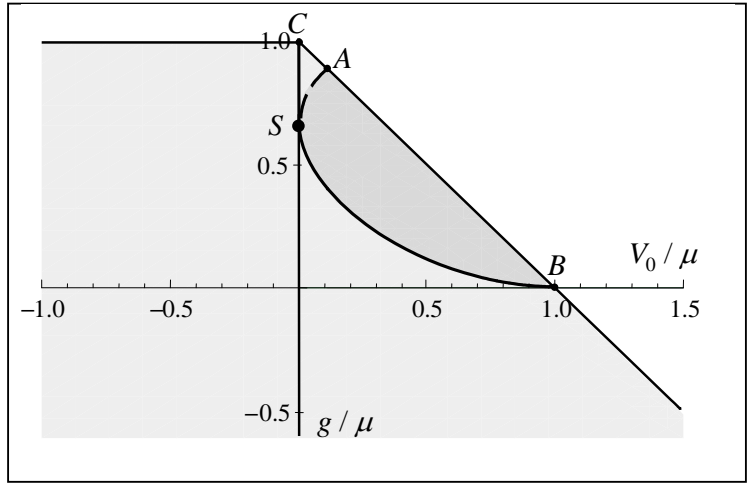

Fig. 1. Transport of a Bose-Einstein condensate above a step potential: in the region outside the curve $C_{CSBAC}$ the solution has an oscillatory asymptote at $x \to +\infty$ while inside $C_{CSBAC}$ it exponentially grows. On the boundary $C_{CSBAC}$ the solution is restricted and non-oscillatory. The singular point $S = \{V_0 = 0, g/\mu = 2/3\}$ is marked by a large filled circle.

In contrast, $e_2$ is positive on the lower branch *SB* of the curve (Fig. 1, solid line *SB*) explicitly given as

$$C_{SB} : \left\{ g = \frac{2}{9} \left( \sqrt{3\mu - 2V_0} - \sqrt{V_0} \right)^2 \right\}. \tag{17}$$

In this case, the solution is non-oscillatory (Fig. 2). It reads

$$p = 1 + e_1 \left( \tanh(k_a x) \right)^2, \quad p(+\infty) = k/\sqrt{g}. \tag{18}$$

If $D > 0$, the elliptic parameters $e_1$ and $e_2$ are real and distinct. In this case the behavior of the solution again depends on whether $\sqrt{ge_2}$ is real or complex (see Eq. (14)). In the latter case, i.e., when $ge_2 < 0$, the solution exponentially diverges. This occurs in the region given by the curved-triangular closed curve $C_{ASCA}$ given as

$$C_{ASCA}: \left\{ V_0 = 0 \, \& \, g = (\mu - V_0) \, \& \, g = \frac{2}{9}\left(\sqrt{3\mu - 2V_0} + \sqrt{V_0}\right)^2 \right\}. \tag{19}$$

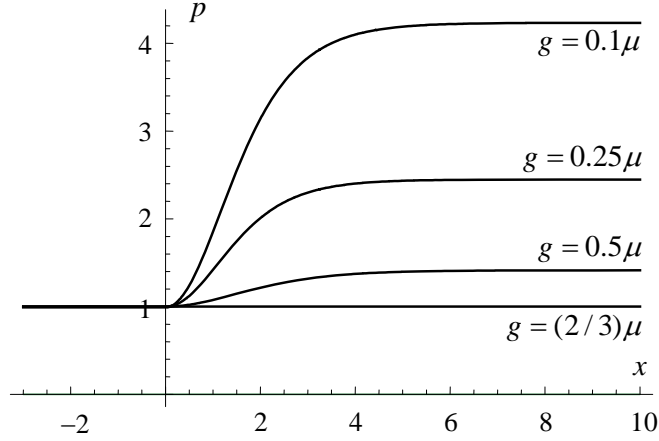

Fig. 2. The non-oscillatory solution on the boundary $C_{SB}$. At $g \to 0$ and, accordingly, $V_0 \to \mu$, the final value $p(+\infty) = k/\sqrt{g}$ increases finally diverging at the point $B = \{V_0 = \mu, g = 0\}$.

In order to understand the reason why this happens, we note that in the particular case when $e_1 = 0$ we encounter a triple root of the cubic polynomial defined by the right-hand side of Eq. (9) so that the equation is rewritten as $(p')^2 = -4g(1-p)^3$. This situation is faced in the point $S = \{V_0 = 0, g/\mu = 2/3\}$, which is thus the most singular point in the parameter space because in this case the solution is $p \equiv 1$, so that $p' = 0$ everywhere and, hence, strictly speaking, Eq. (9) can be applicable at this point only as a limit. Since one initially would expect that the solution on the upper branch of the elliptic curve $C_{ASB}$ would be similar to that on the lower one this point makes a change. The dynamics of the system changes from one evolution scenario to another one. It is understood that this change is caused by the circumstance that at this singular point the elliptic curve $C_{ASB}$ touches the vertical line $V_0 = 0$ corresponding to the case when no barrier is present and, hence, $p(x) \equiv 1$ everywhere along that line.

Finally, we note that everywhere outside the closed curve $C_{CSBAC}$ $ge_2 > 0$ and Eq. (14) defines a periodic function with the period given as

$$T = \frac{\pi}{\sqrt{g e_2}} {}_2F_1(1/2, 1/2; 1; m). \qquad (20)$$

The behavior of the solution for $g/\mu = 1/3$ is shown in Fig. 3.

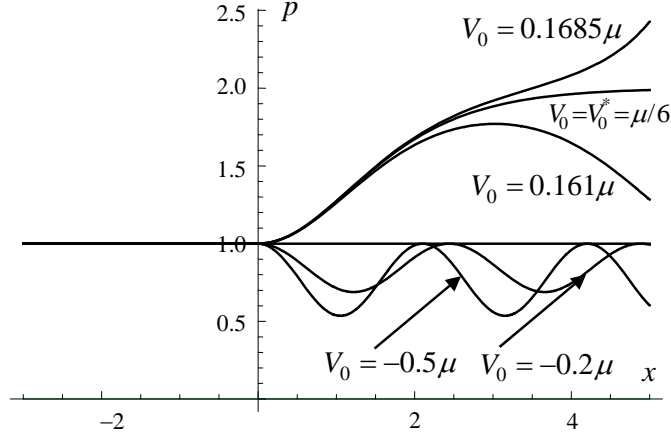

Fig. 3. Evolution of the system at $g/\mu = 1/3$. The solution is oscillatory up to the point $V_0^* = \mu/6$ belonging to the boundary curve $C_{SB}$. At $V_0 = V_0^*$ the solution is non-oscillatory, and for $V_0 > V_0^*$ the solution diverges.

Thus, we have presented a comprehensive analysis of the matter-wave transport above a step potential within the framework of the cubic-nonlinear Schrödinger equation. Using the exact solution of the problem written in terms of the Jacobi elliptic *sn*-function, we have shown that in different regions of the 2D space of involved dimensionless parameters, the nonlinearity and the reflecting potential's height/depth, qualitatively distinct types of wave propagation (oscillatory, non-oscillatory and diverging) are observed. We have shown that the region of the space where the solution is not restricted is defined by an elliptic curve and two lines. We have determined the explicit form of the elliptic curve and have shown that there exists a singular point belonging to this curve that causes a jump from one possible evolution scenario to another one. We have determined the position of this point and have presented in detail the peculiarities of the dynamics of the system for all the regions of involved parameters.

**Acknowledgments**

This research has been conducted in the scope of the International Associated Laboratory IRMAS. The work was supported by the Armenian National Science and Education Fund (ANSEF Grants No. 2464 and No. 2591), Armenian State Committee of Science (SCS Grants No. 11A-1c061 and No. 11RB-026), and Russian Foundation for Basic Research (RFBR Grant No. 09-08-00722). H.A. Ishkhanyan acknowledges SPIE for a 2011 Scholarship in Optics and Photonics.